\documentclass[universe,article,accept,oneauthor,pdftex]{mdpi}

\firstpage{1}
\pubvolume{5}
\issuenum{5}
\articlenumber{126}
\pubyear{2019}
\copyrightyear{2019}
\history{Received: 19 April 2019; Accepted: 17 May 2019; Published: 23 May 2019}
\updates{yes} 

\Title{Influence of Backside Energy Leakages from Hadronic Calorimeters on Fluctuation Measures in Relativistic Heavy-Ion Collisions}

\Author{{Andrey Seryakov}\href{https://orcid.org/0000-0002-5759-5485}{\orcidicon}}

\AuthorNames{Andrey Seryakov}

\address[1]{Laboratory of Ultra-High Energy Physics, St. Petersburg State University, {Saint Petersburg 199034, Russia}; a.seryakov@spbu.ru or seryakov@yahoo.com
}

\corres{Correspondence: seryakov@yahoo.com}

\abstract{\textls[-16]{The phase diagram of the strongly interacting matter is the main research subject for different current and future experiments in high-energy physics. System size and energy scan programs aim to find a possible critical point. One of such programs was accomplished by the fixed-target NA61/SHINE experiment in 2018. It includes six beam energies and six colliding systems: p + p, Be + Be, Ar + Sc, \mbox{Xe + La}, Pb + Pb and p + Pb. In this study, we discuss how the efficiency of centrality selection by forward spectators  influences multiplicity and fluctuation measures and how this influence depends on the size of colliding systems. We use {SHIELD and EPOS} Monte-Carlo (MC) generators along with the wounded nucleon model, introduce a probability to lose a forward spectator and spectator energy loss. We show that for light colliding systems such as Be or Li even a small inefficiency in centrality selection has a dramatic impact on multiplicity scaled variance. Conversely, heavy systems such as Ar + Sc are much less prone to the effect.}}

\begin{document}

\section{Introduction}
Fluctuation measures are considered to be an important tool in the search of the possible critical point of the strongly interacting matter. However, fluctuation quantities are sensitive to various effects along with the critical behavior \cite{Luo,Vovchenko,NuXu,MandP} such as volume fluctuations \cite{VolumeAndAuto,StronglyIntensive}, resonance decays \cite{resonance}, beam and target material impurities \cite{machiek} and detector inefficiencies.

Experiments in relativistic heavy ion collisions use different techniques to reduce volume fluctuations by selecting centrality classes. The procedure aims to select events with a restricted number of particle production sources or volume. The centrality selection may be accomplished by measuring produced particle multiplicity in a specific rapidity interval along with energy of non-interacted nucleons-spectators by forward hadronic calorimeters. Although the multiplicity based approach introduces, a bias on any  fluctuations due to correlations between multiplicities even in different acceptance windows. However, it~is worth noting that this bias can be well reproduced and estimated by using MC generators.

Contrary to this, a solely spectators based centrality selection provides an unbiased method to restrict the collision volume. Technically, it can be accomplished only in fixed target experiments, as it is possible there to place a hadronic calorimeter exactly at the beam line. Nevertheless, such calorimeters suffer from hadronic shower energy leakages from the surface and have much lower resolution capabilities compared to multiplicity detectors. In this paper, we study an influence of the energy leakage from the calorimeter back surface on the average multiplicities and the multiplicity scaled variance and its dependency on the colliding system size.

\section{Study with a Geant Calorimeter Model} \label{sec2}
\begin{figure}[H]
\centering
\includegraphics[width=8 cm]{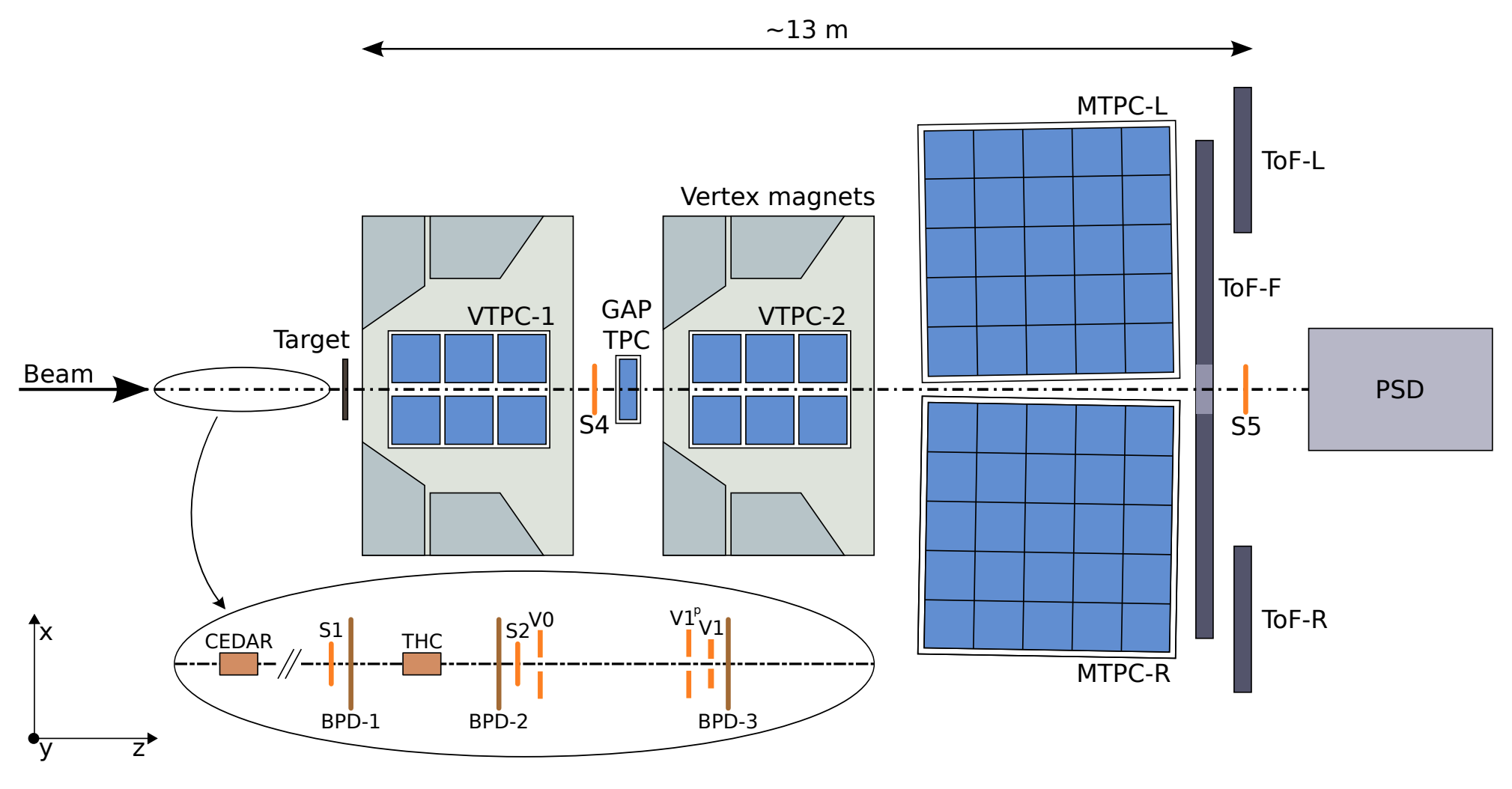}
\includegraphics[width=5.5 cm]{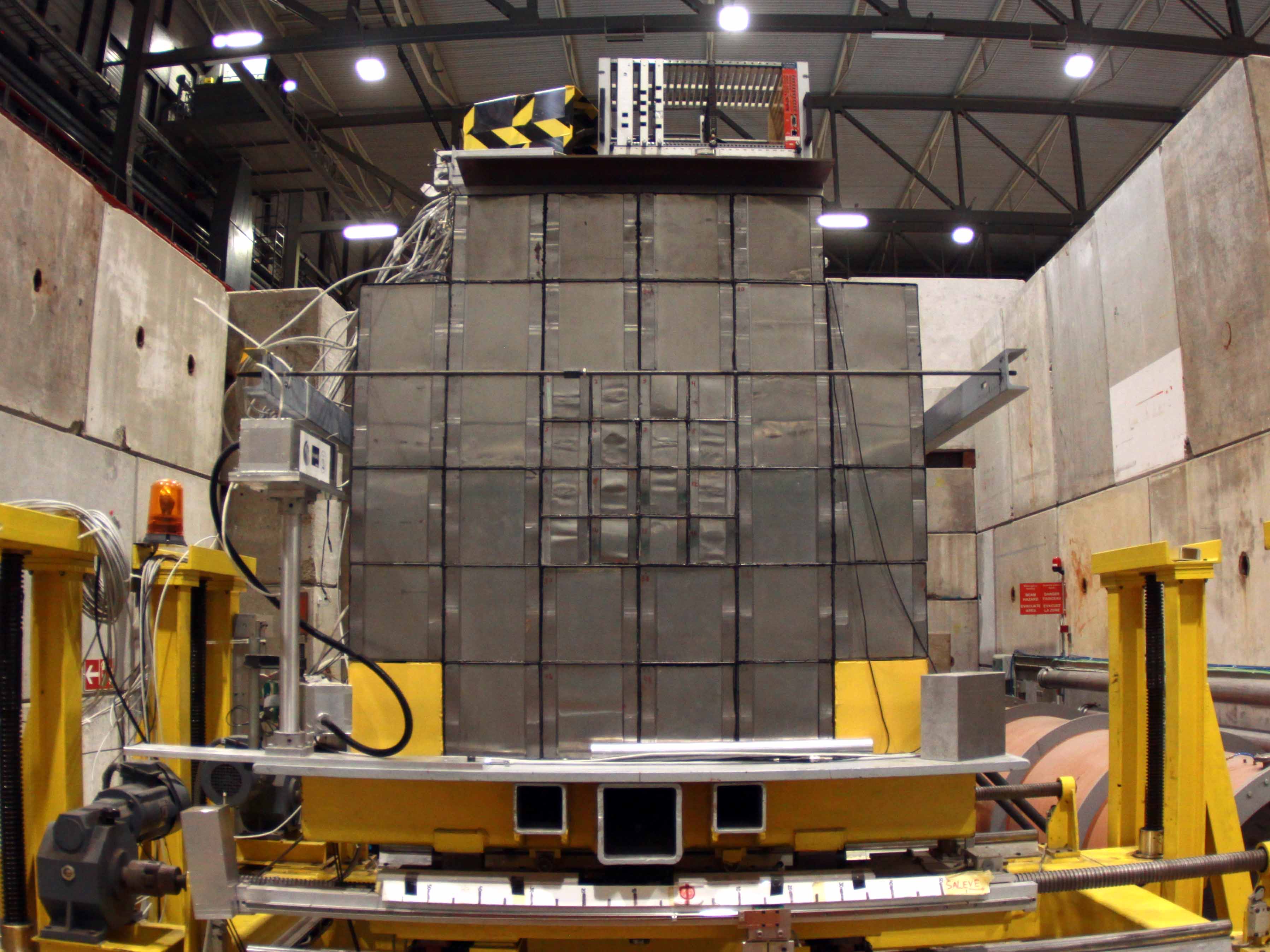}
\caption{A scheme of the NA61/SHINE experiment \cite{facility} and a front view photo of the Projectile Hadronic Calorimeter. PSD is used for triggering, centrality and event plane determination.} \label{fig1}
\end{figure}

The main motivation for this work was a study of how Projectile Spectator Detector (PSD) \cite{Marina} influences the measured quantities in the NA61/SHINE collaboration \cite{facility}. PSD is a segmented modular hadronic calorimeter, which is used for triggering, centrality and event plane determination. The detector consists of 44 independent modules and each of them has 60 lead (16 mm) + scintillator (4 mm) layers. The~total length of PSD is about 1.2 m, which corresponds to approximately 5.6 interaction lengths. A~scheme of the NA61/SHINE setup and a photo of PSD are presented in  Figure \ref{fig1}.

\begin{figure}[H]
\centering
\includegraphics[width=5 cm]{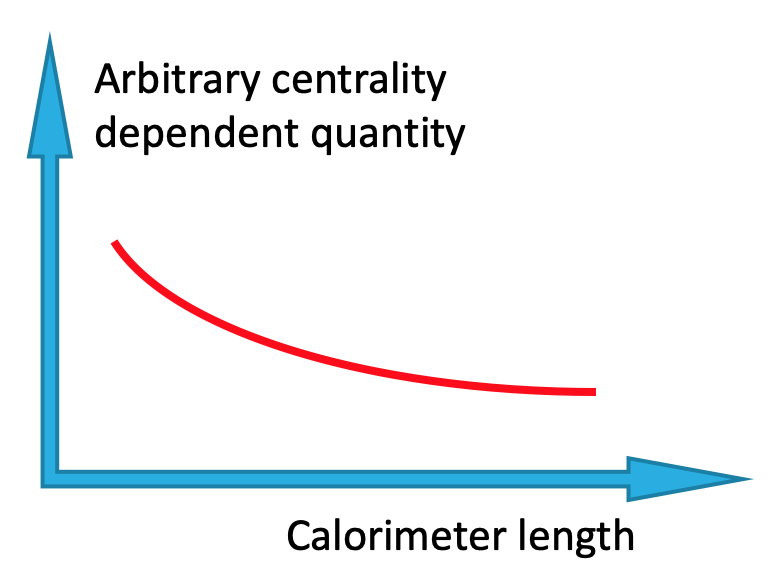}
\caption{A  {sketch} of how a collision volume depended quantity behaves with the increase of a centrality calorimeter length.} \label{fig2}
\end{figure}

Signals from 60 scintillators in each module are grouped by six, therefore, the design of PSD allows   collecting information from ten independent areas along the beam axis inside the calorimeter. This makes it possible to study dependencies of different quantities on the calorimeter lengths by selecting centrality by a reduced number of scintillator groups. It is expected that any centrality sensitive measure will saturate at one point with an increase of calorimeter length (see  Figure \ref{fig2}). On such plot, the 0 limit corresponds to 0 centrality detector efficiency, an absence of centrality selection and to minimum bias events. The~right limit is an absence of any energy leakage from the hadronic calorimeter backside.
\vspace{-6pt}

\begin{figure}[H]
\centering
\includegraphics[width=5 cm]{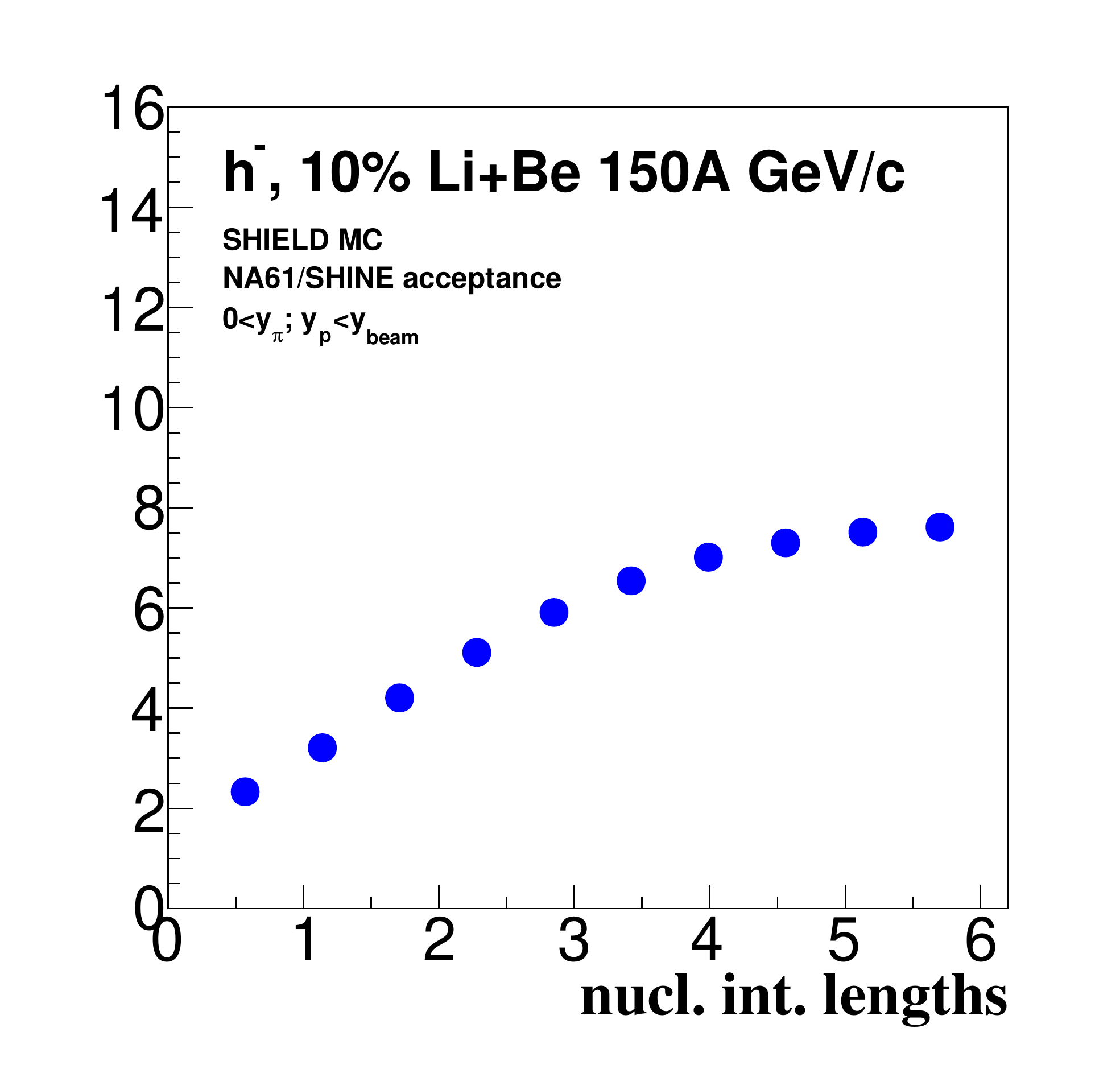}
\includegraphics[width=5 cm]{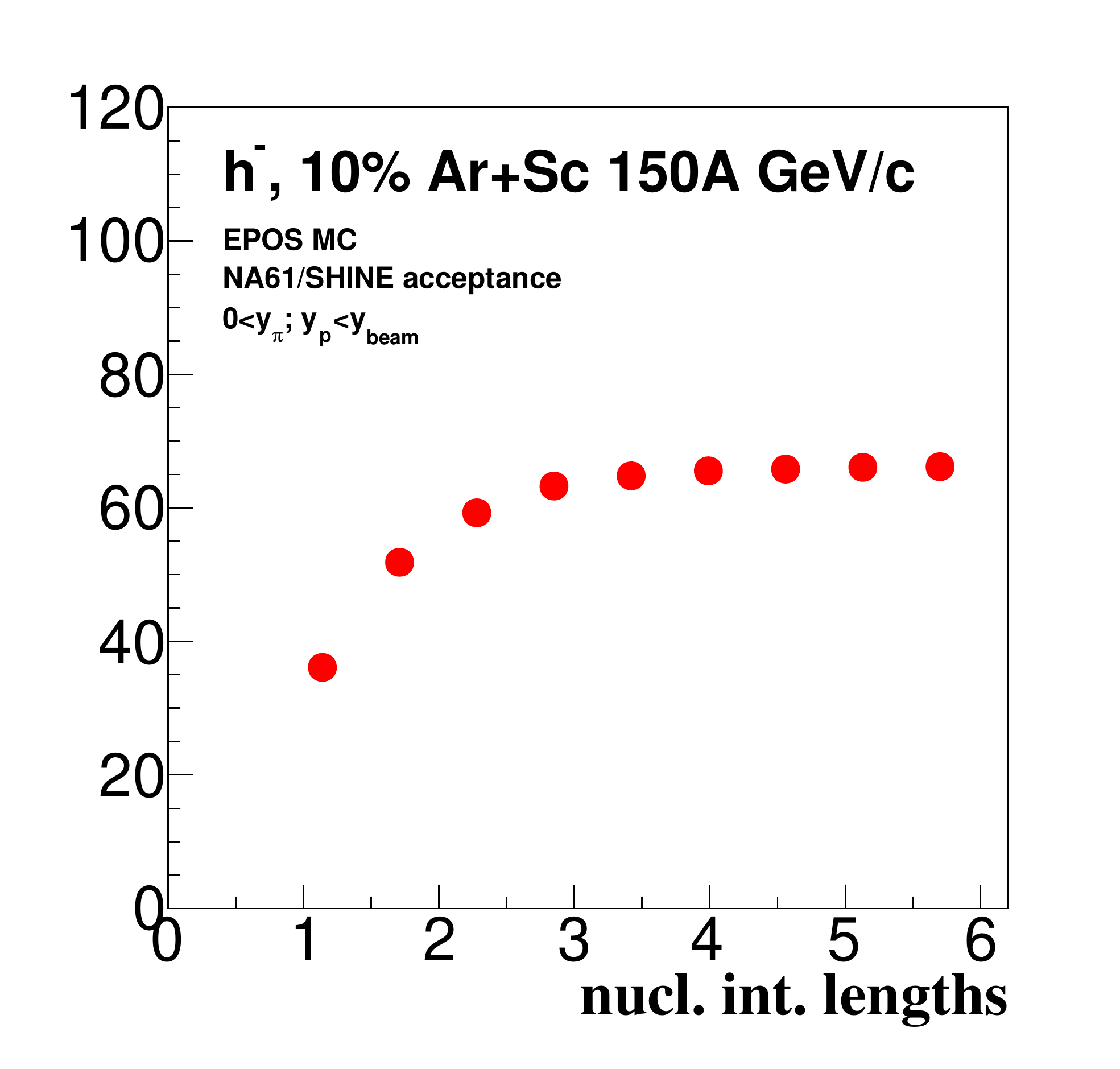}
\includegraphics[width=5 cm]{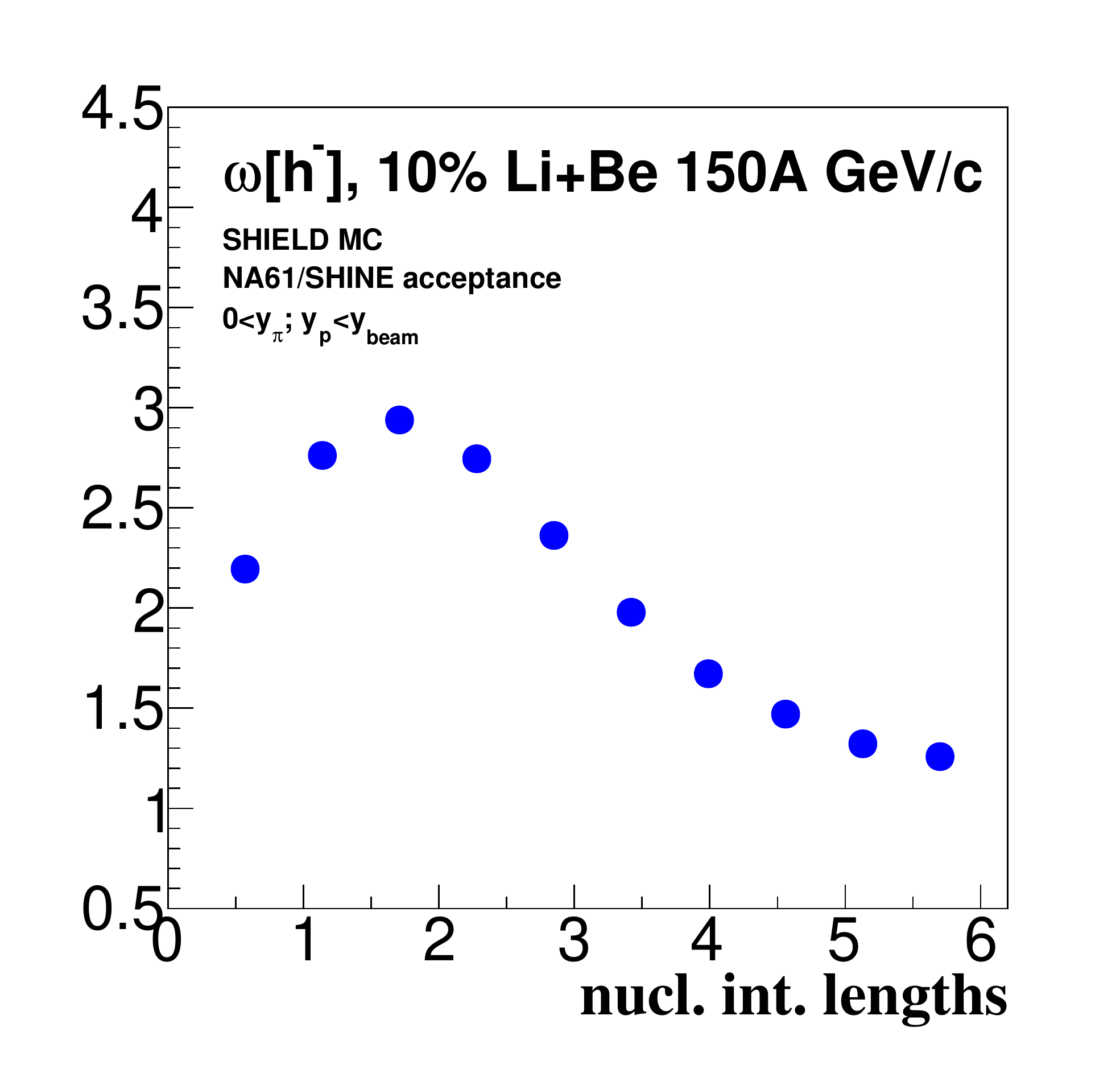}
\includegraphics[width=5 cm]{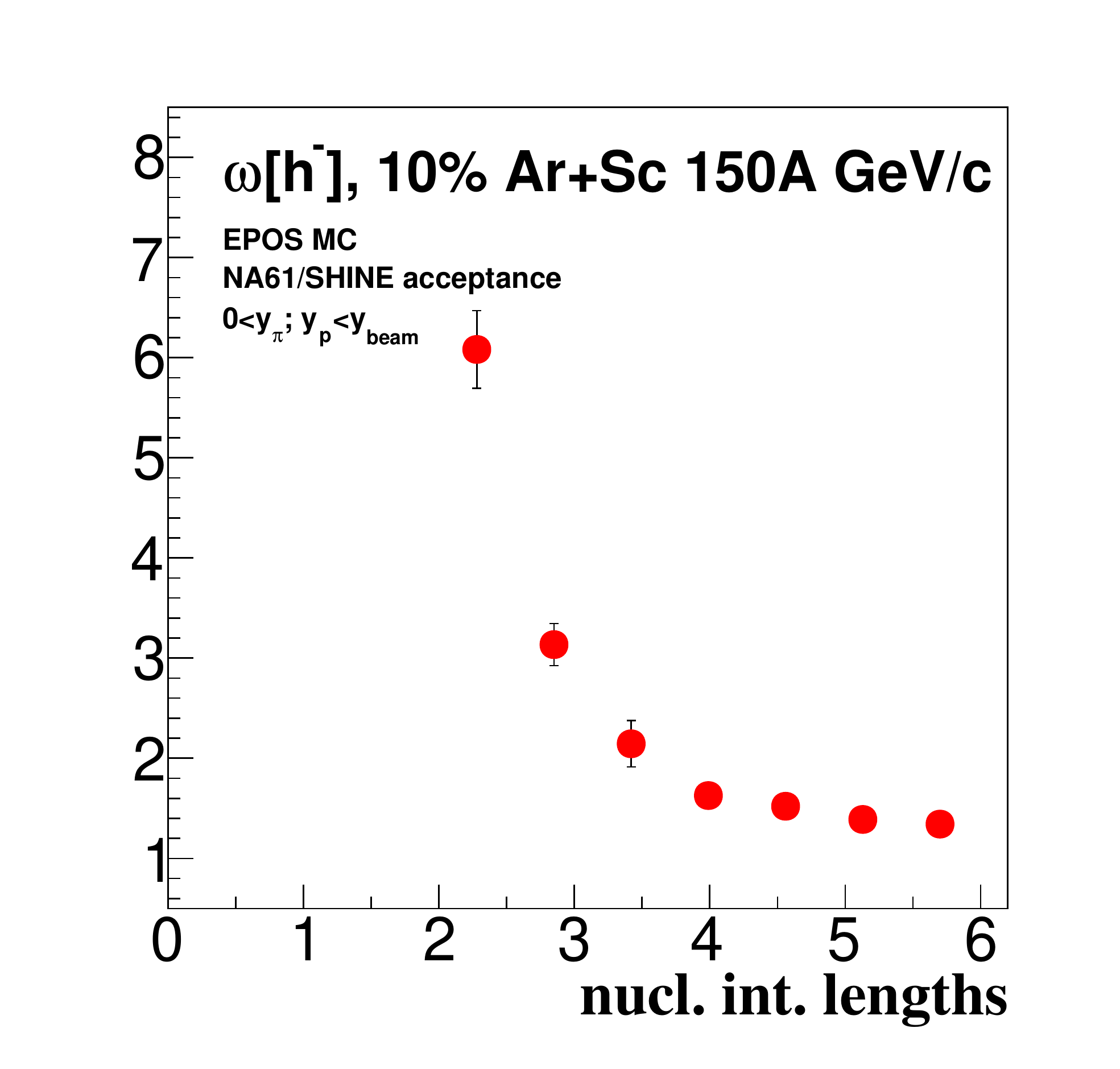}
\includegraphics[width=5 cm]{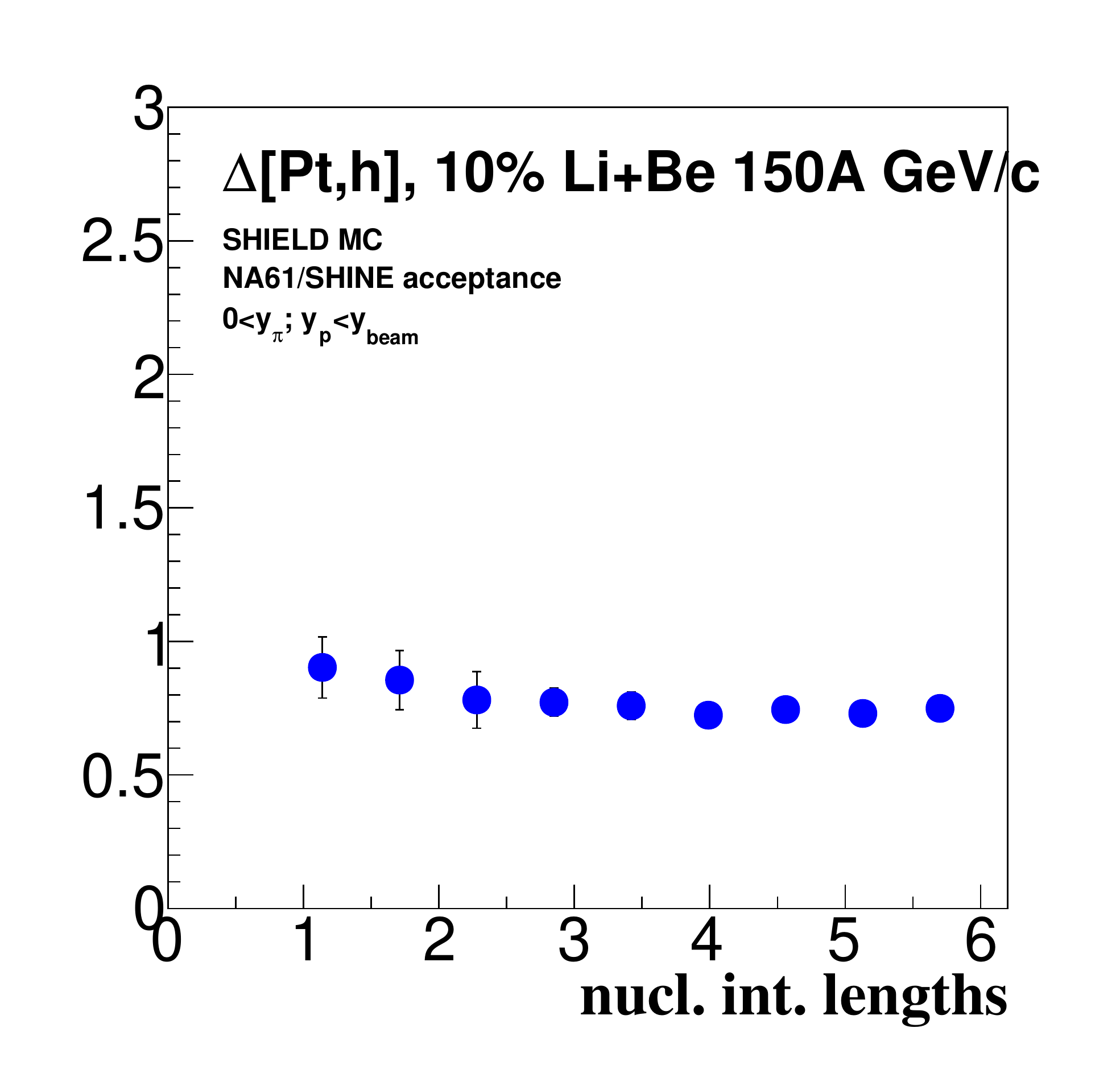}
\includegraphics[width=5 cm]{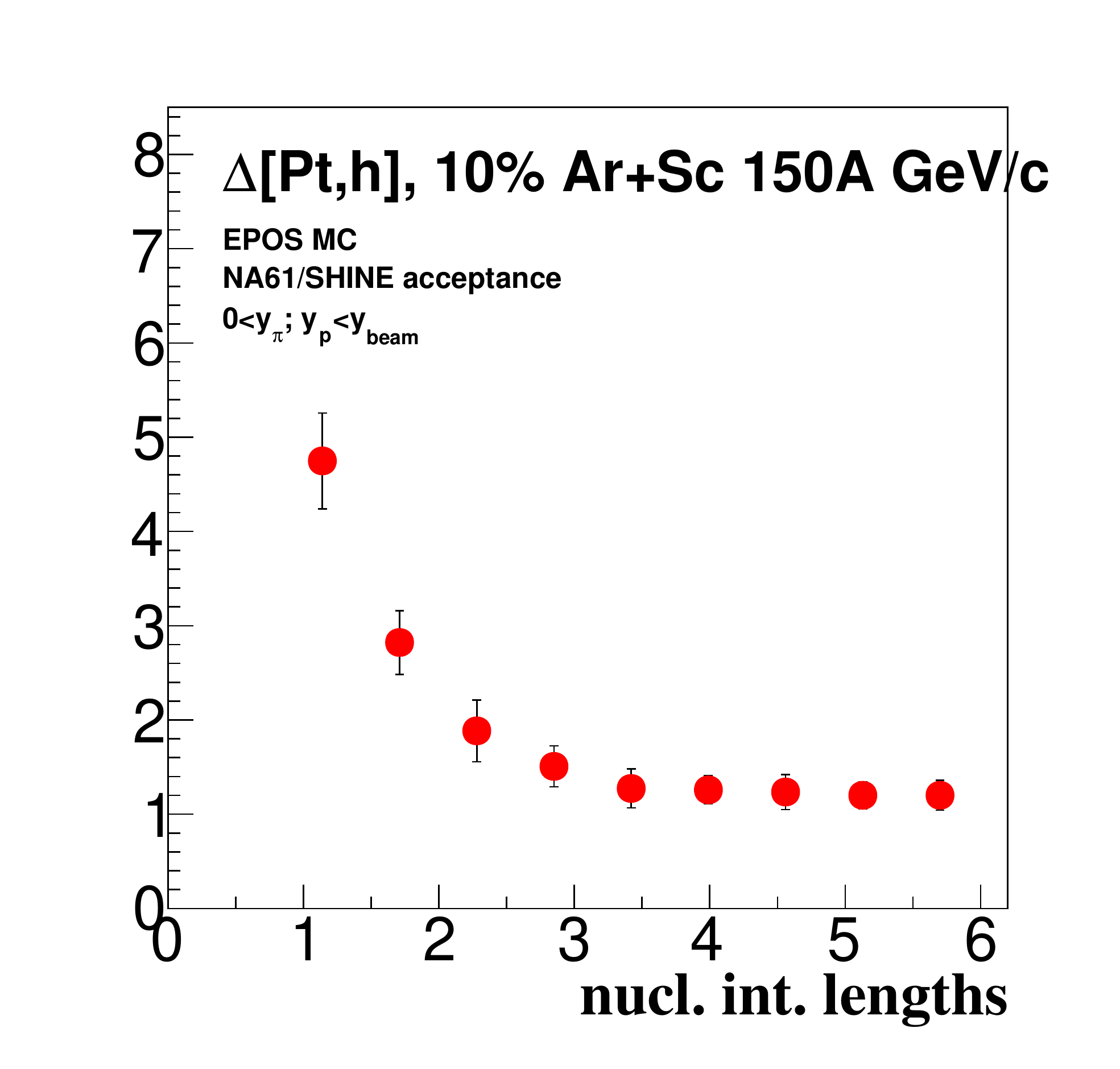}
\includegraphics[width=5 cm]{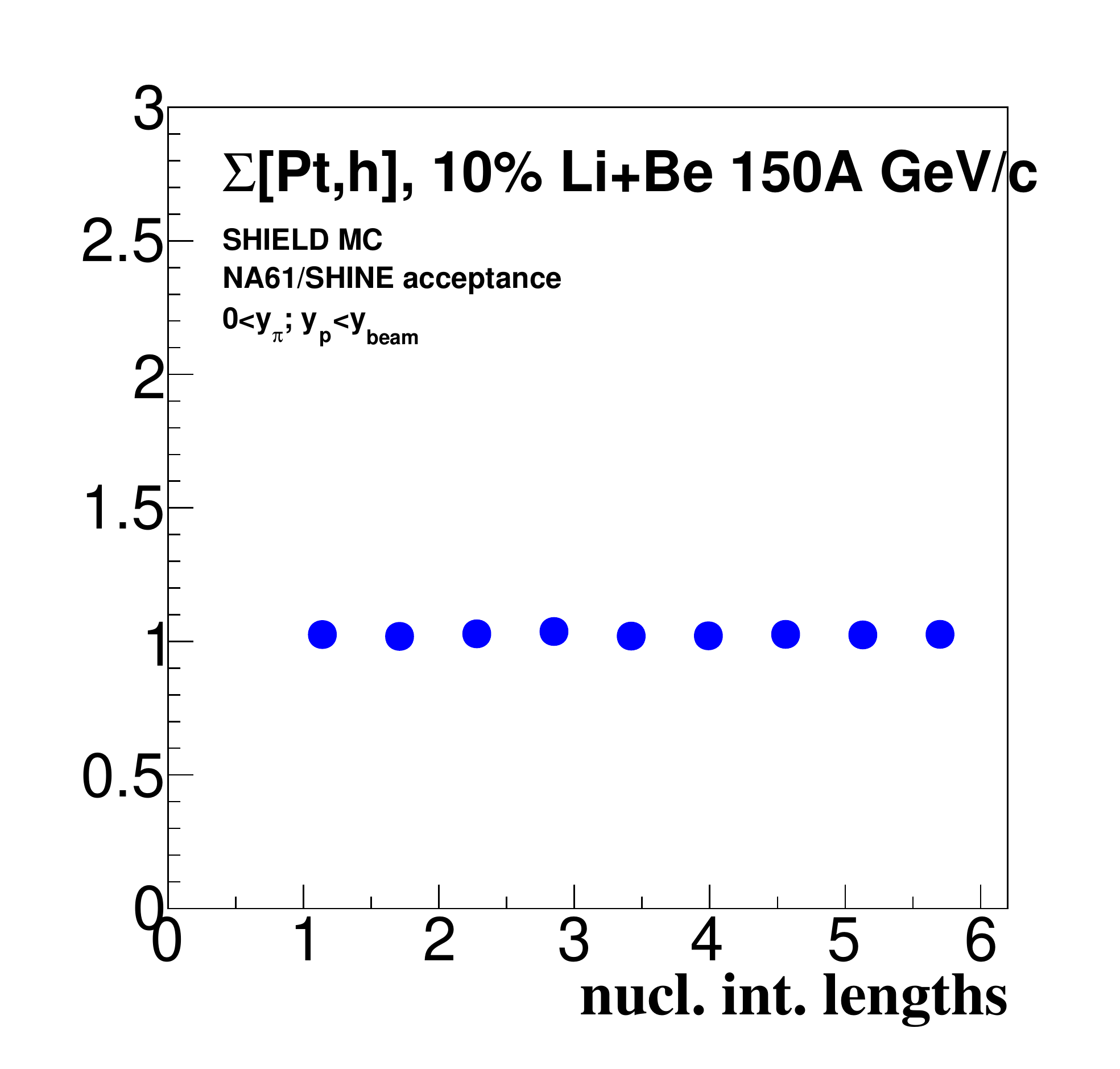}
\includegraphics[width=5 cm]{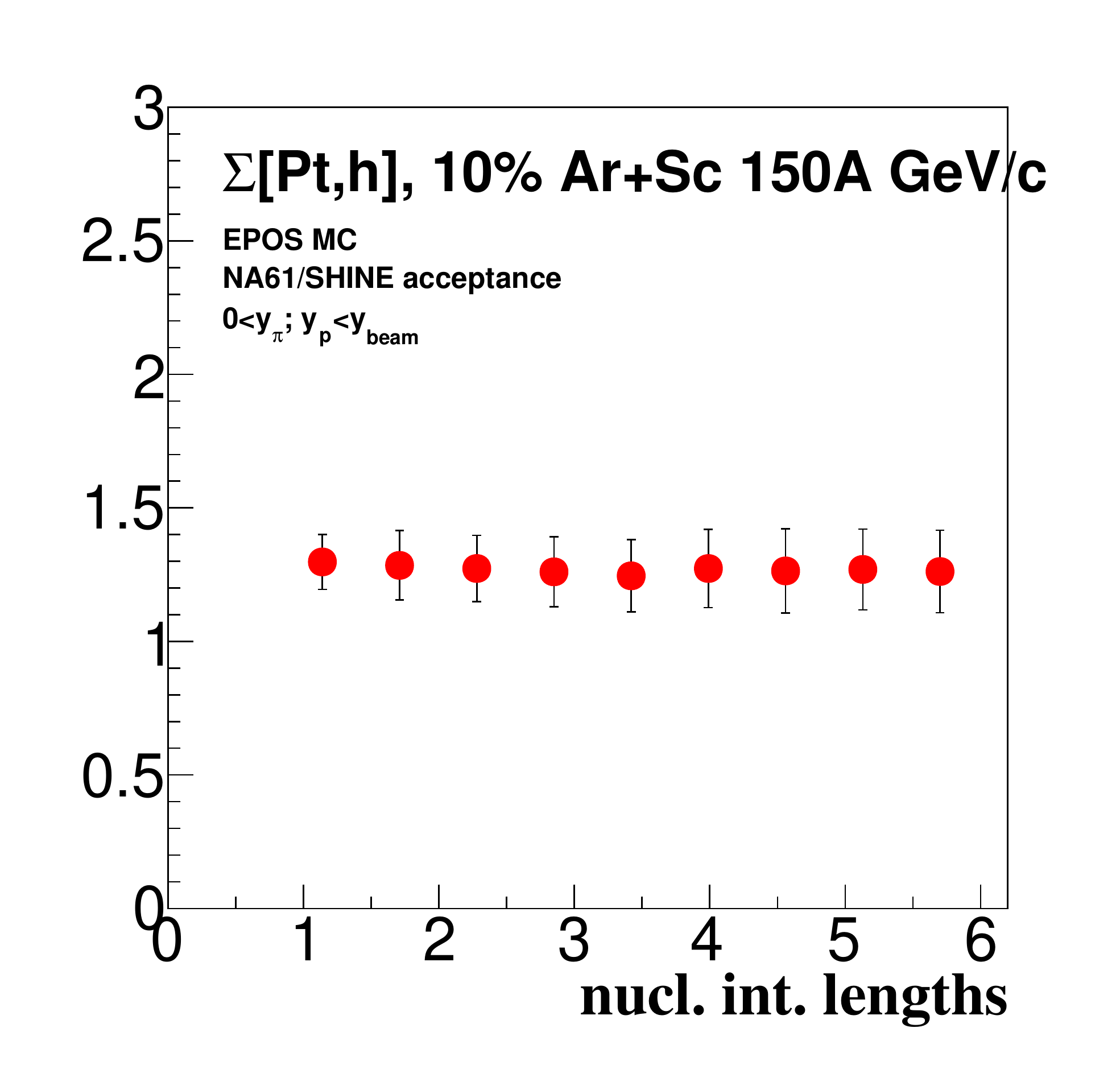}
\includegraphics[width=5 cm]{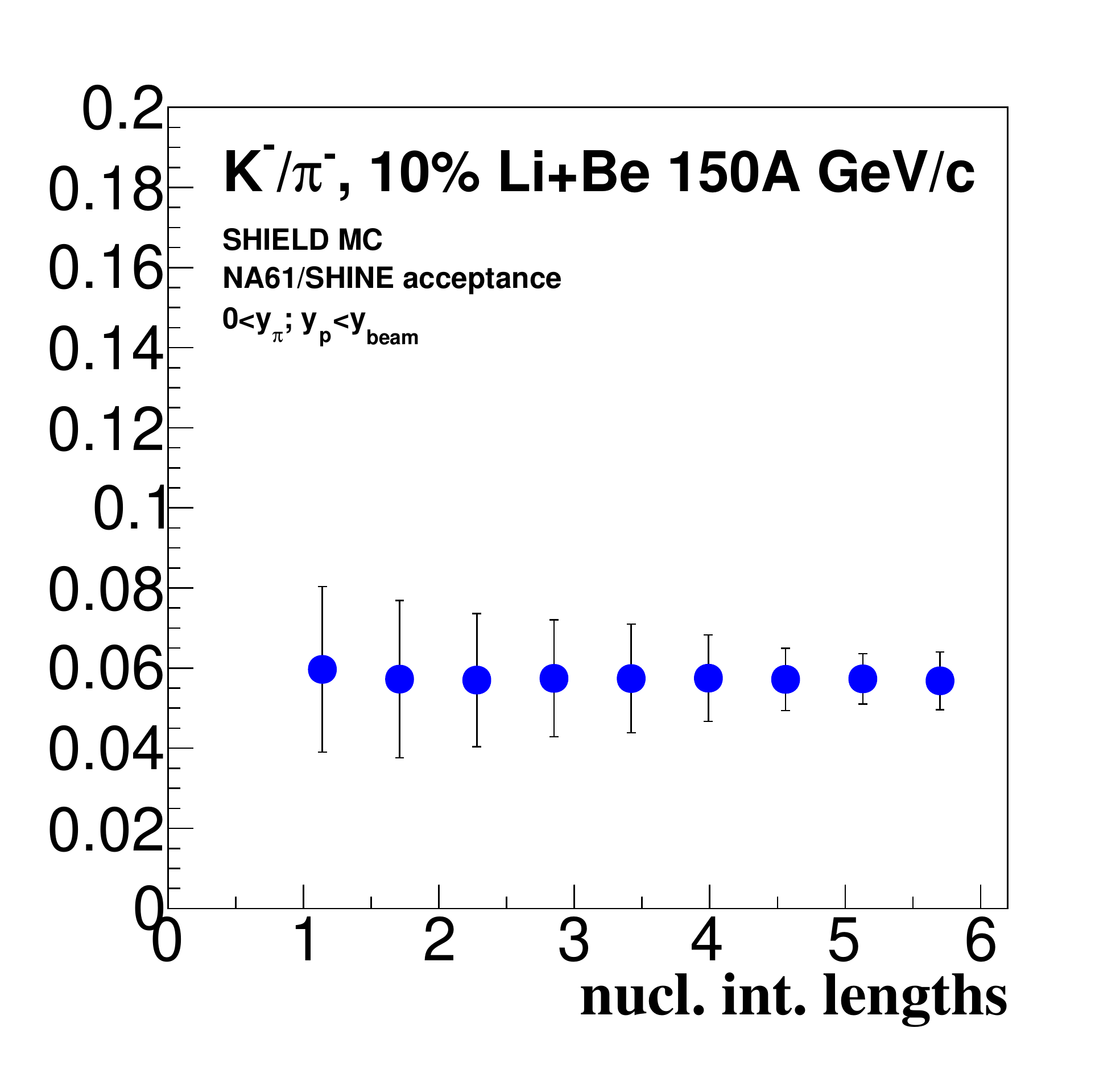}
\includegraphics[width=5 cm]{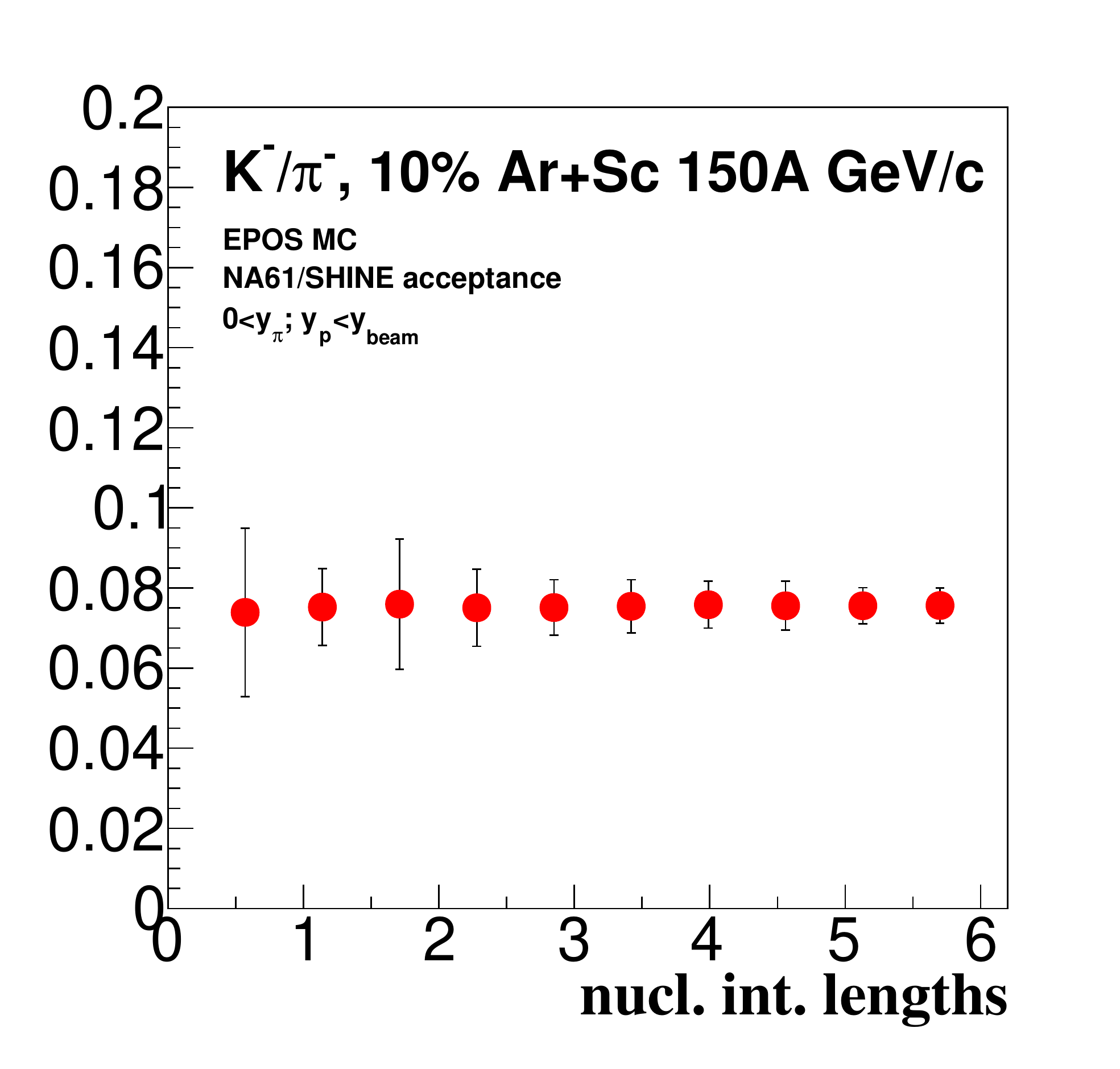}
\caption{Comparison of different measures behavior versus the centrality calorimeter length for \mbox{Li + Be} (blue dots) and Ar + Sc (red dots) 150\textit{A} GeV/c collisions. The first two plots present an average number of negatively charged hadrons, the second two show the negative charged hadrons scaled variance, next~four show two strongly-intensive quantities $\Delta$[Pt,h] and $\Sigma$[Pt,h] \cite{StronglyIntensive2} and the last two show a ratio of the average number of negative charged kaons to the average number of negative charged pions. All results were calculated in the NA61/SHINE acceptance \cite{acceptance}.} \label{fig3}
\end{figure}

Two MC datasets were generated with the GEANT4 \cite{GEANT} PSD simulation for studying the energy leakage influence on different systems: 100,000 events of 150\textit{A} GeV/c ${}^7$Li + ${}^9$Be SHIELD MC \cite{SHIELD} and 40,000 events of 150\textit{A} GeV/c  ${}^{40}$Ar + ${}^{45}$Sc EPOS 1.99 MC \cite{EPOS}. ${}^7$Li was chosen instead of experimentally used ${}^7$Be as the first one is stable and can be simulated by SHIELD MC. It was possible to compare two completely different MC generators as the studied effect was purely detector based.  Moreover, it does not depend on the spectator transverse characteristics as we studied longitudinal shower propagation that is insensitive to a hit position. In each dataset, we selected centrality on the different length of the detector from $\approx$1.1 to 5.6 interaction lengths. The results for average multiplicities, multiplicity ratios and fluctuation quantities are presented in Figure \ref{fig3} for  $^7$Li + $^9$Be   and for  $ {}^{40}$Ar + ${}^{45}$Sc collisions.

Two main conclusions may be drawn from the results:
\begin{itemize}[leftmargin=*,labelsep=5.8mm]
\item The 5.6 interaction lengths were not enough to eliminate the influence of backside energy leakage in light colliding systems (${}^7$Li + ${}^9$Be) on volume fluctuations as the dependencies did not saturate. The middle size systems as ${}^{40}$Ar + ${}^{45}$Sc were much less prone to the effect. However, Ar + Sc data are more sensitive to energy leakage in case of a short calorimeter.
\item Mean multiplicities, scaled variance $\omega$[h] = (<h$^2$> $-$ <h>$^2$)/<h> and strongly intensive  $\Delta$[Pt,h]~\cite{StronglyIntensive2} were sensitive to the effect, while mean multiplicity ratios and   another strongly intensive quantity~$\Sigma$[Pt,h] showed steady behavior. The instability of $\Delta$[Pt,h] contradicted the   presumption that~such quantities do not depend on the volume fluctuations. Therefore, it was clear that assumptions which lead to the construction of the strongly intensive {measure} {$\Delta$[Pt,h]} \cite{StronglyIntensive2} are not fulfilled even in MC generators. Investigation of other measures sensitivities transcends the scope of this work.
\end{itemize} 

\section{Study within a Wounded Nucleon Model}\label{sec3}

{A simple wounded} nucleon model (WNM) was created to understand the unexpected sensitivity of light systems to the energy leak. Three different colliding systems were considered:  \mbox{${}^7$Li + ${}^9$Be}, \mbox{${}^{35}$Cl + ${}^{40}$Ca} and \mbox{${}^{208}$Pb + ${}^{208}$Pb} with 150 \textit{A} GeV/c beam momentum (\begin{math} \sqrt{S_{NN}}\approx\end{math} 17\textit{A} GeV). Nucleon density profiles were taken from \cite{table}. Nucleon core effect was not taken into account. Alpha clustering was   not implemented as the goal of the study was to check how the sensitivity to the energy leakage depends on the number of nucleons in colliding systems. Inelastic nucleon--nucleon cross-section was taken equal to $31.75$ mb. Multiplicity was introduced based on the number of wounded nucleons; in other words, each wounded nucleon produced a random number of charged particles, which were distributed according to a Poisson with \begin{math} <N_{ch}>  = 3.5\end{math}.

\begin{figure}[H]
\centering
\includegraphics[width=16 cm]{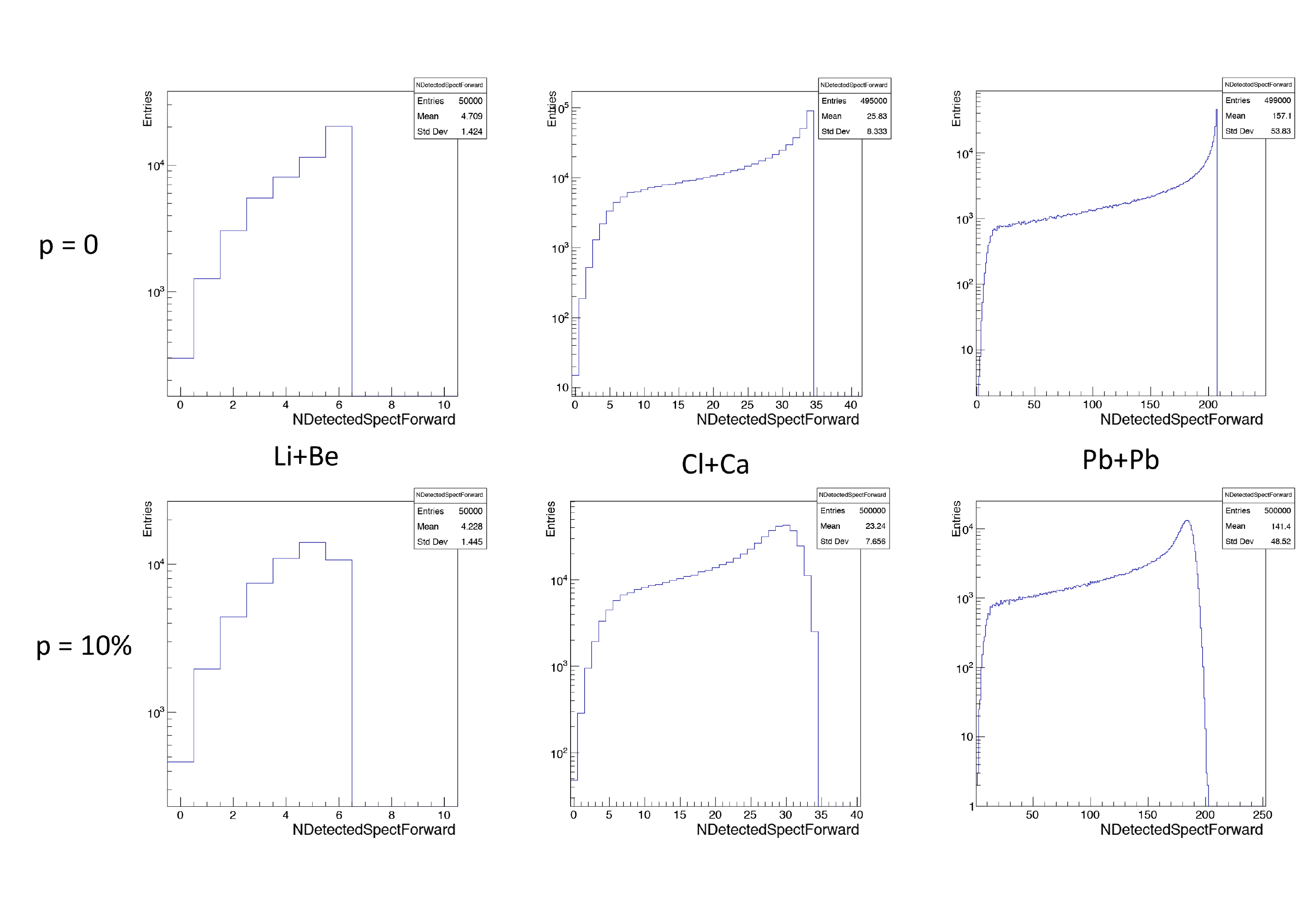}
\caption{Detected forward spectators distributions for Li + Be, Cl + Ca and Pb + Pb in WNM with a probability to loss a nucleon of 0\% and 10\%.} \label{fig4}
\end{figure}

In the first version of the model, we introduce dcentrality selection based on the number of forward nucleon-spectators and a probability to lose each of them \begin{math}p\end{math}. Distributions of forward nucleon spectators for \begin{math}p=0\%\end{math} and \begin{math}10\%\end{math} are shown in  Figure \ref{fig4}.

The 10\% of events with a lower number of detected forward nucleons spectators were selected as the most central ones. If the boundary between classes di   not coincide with the boundary between integer numbers of forward spectators N, then a fraction of events with N + 1 forward spectators was taken to obtain exactly 10\% of the whole data sample.

\begin{figure}[H]
\centering
\includegraphics[width=4 cm]{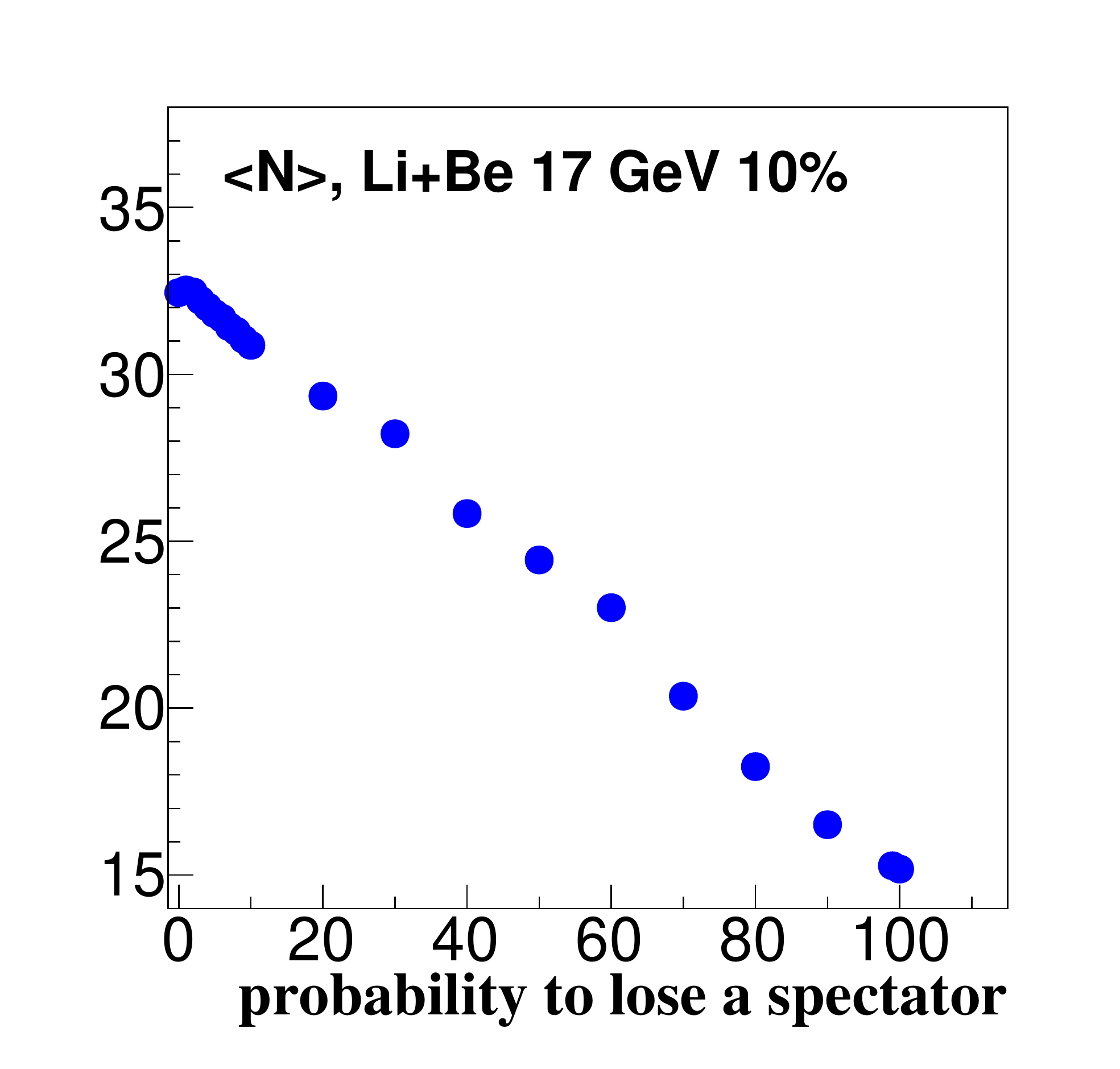}
\includegraphics[width=4 cm]{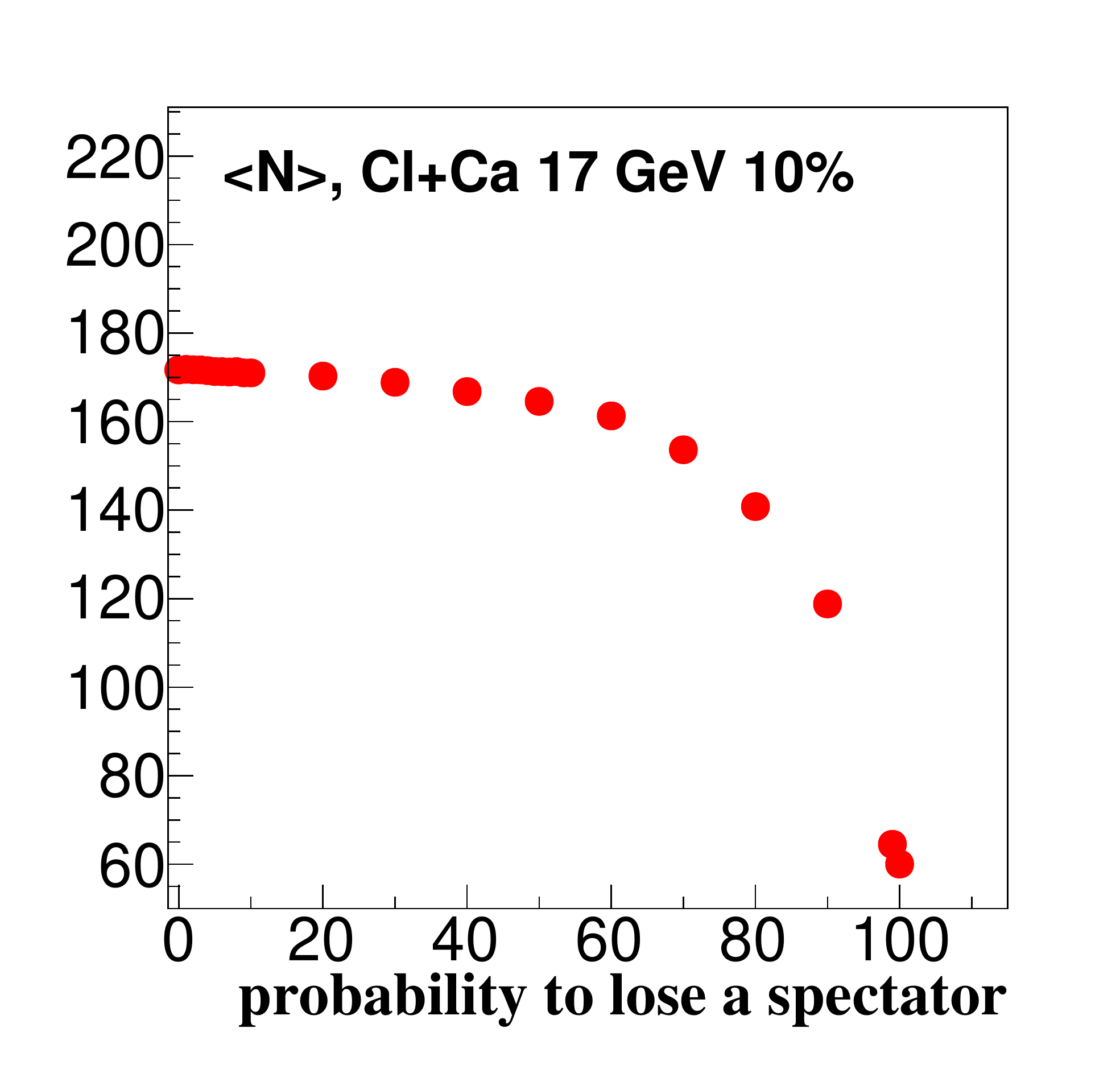}
\includegraphics[width=4 cm]{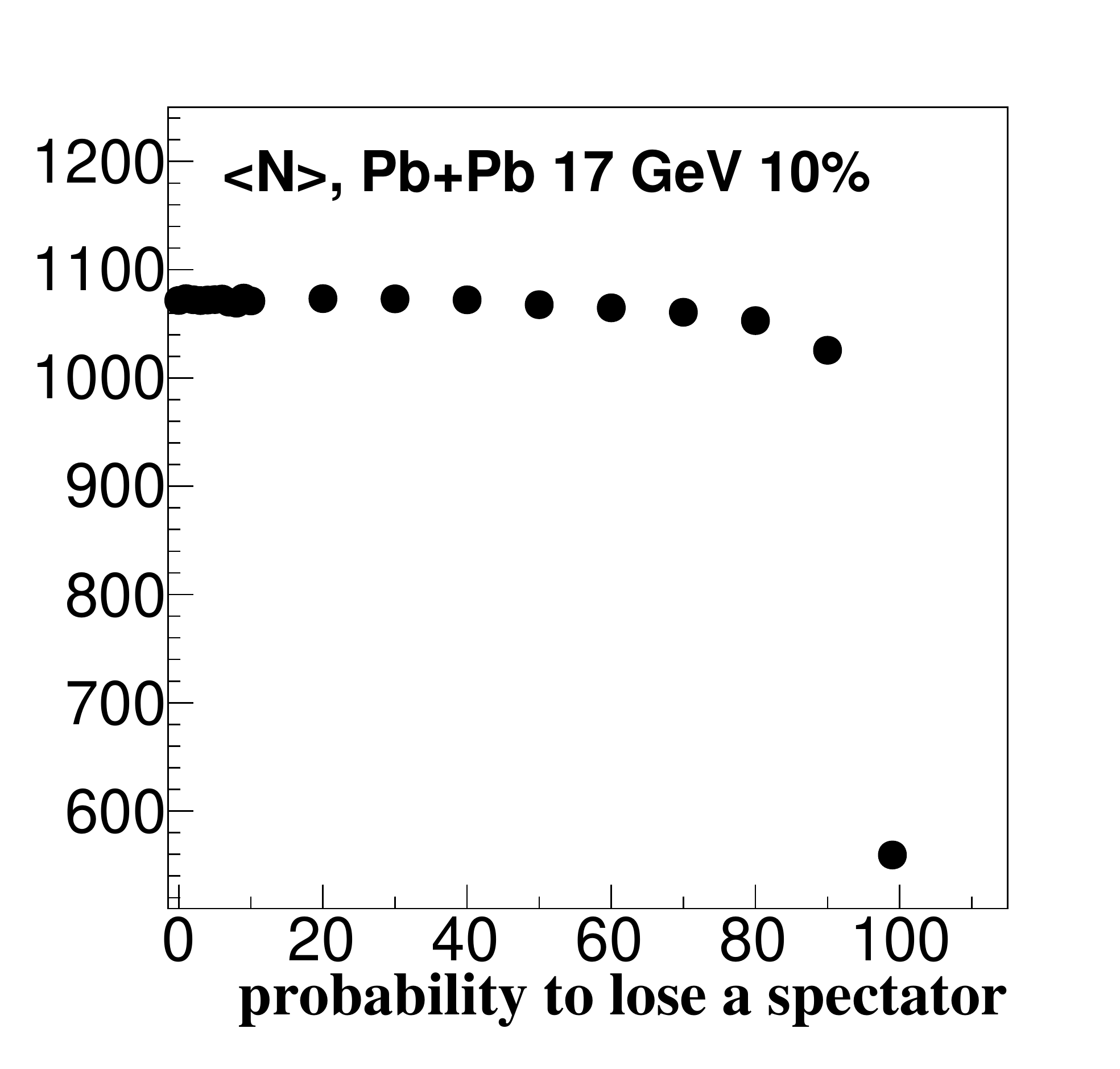}
\includegraphics[width=4 cm]{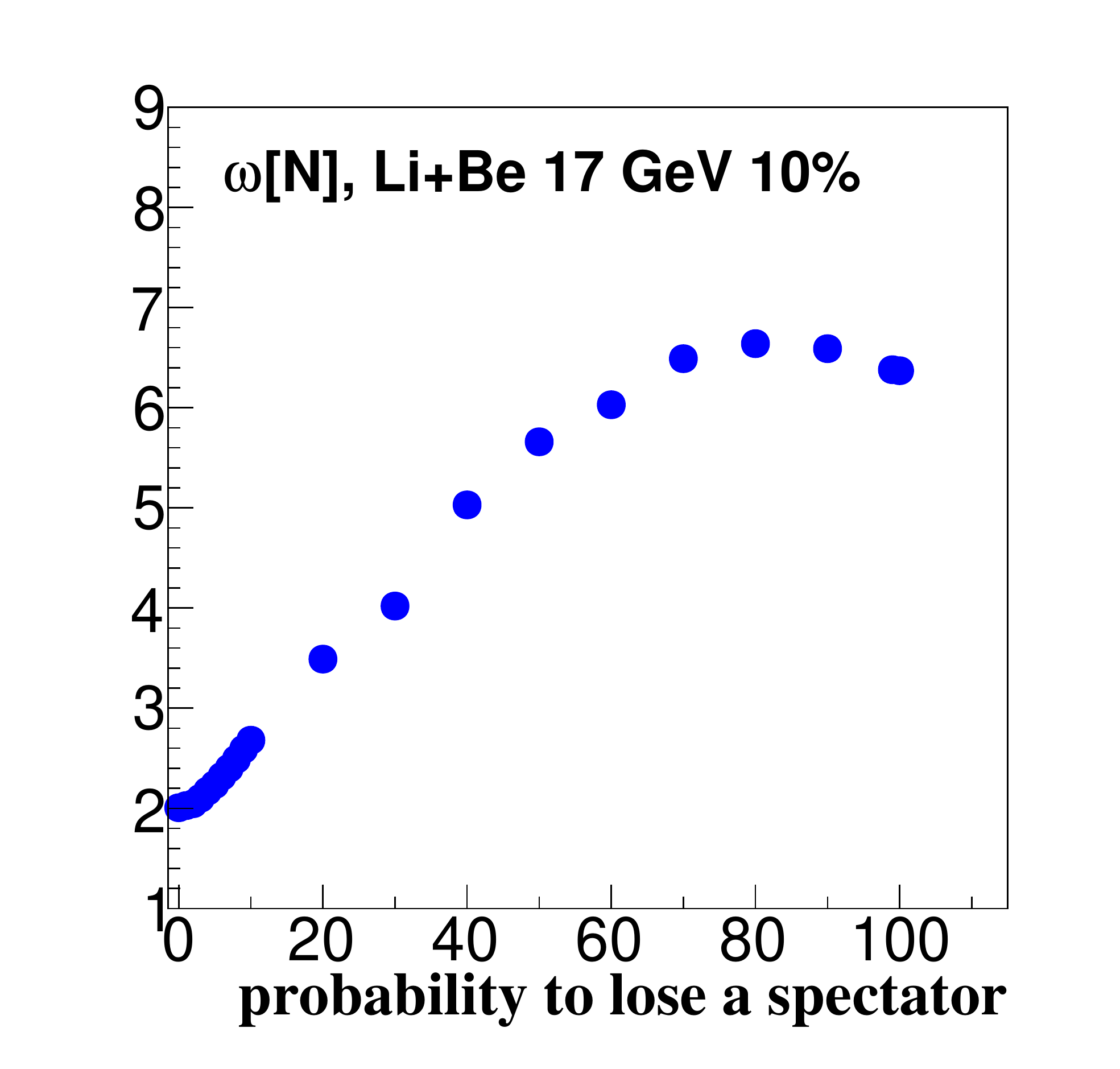}
\includegraphics[width=4 cm]{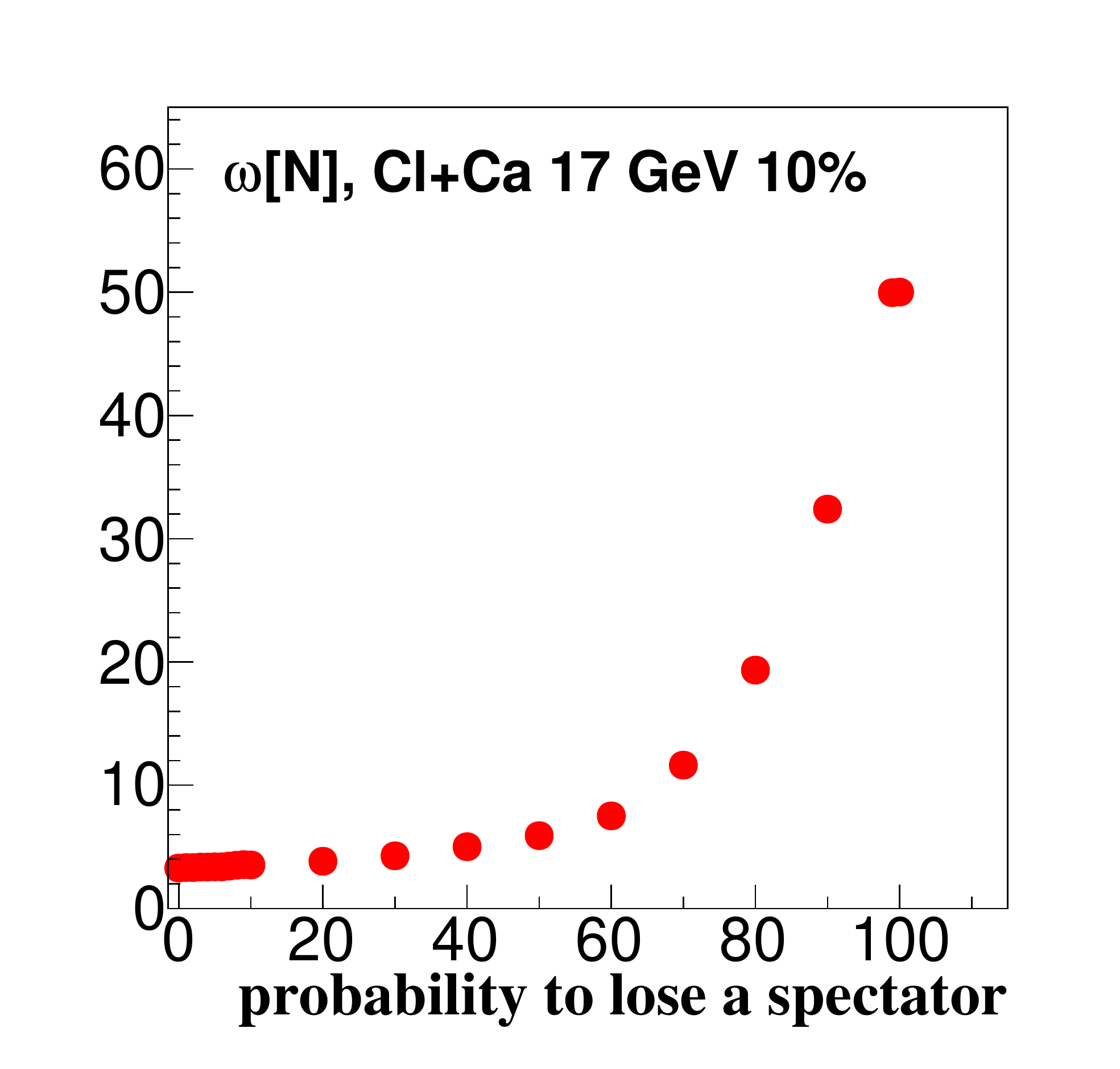}
\includegraphics[width=4 cm]{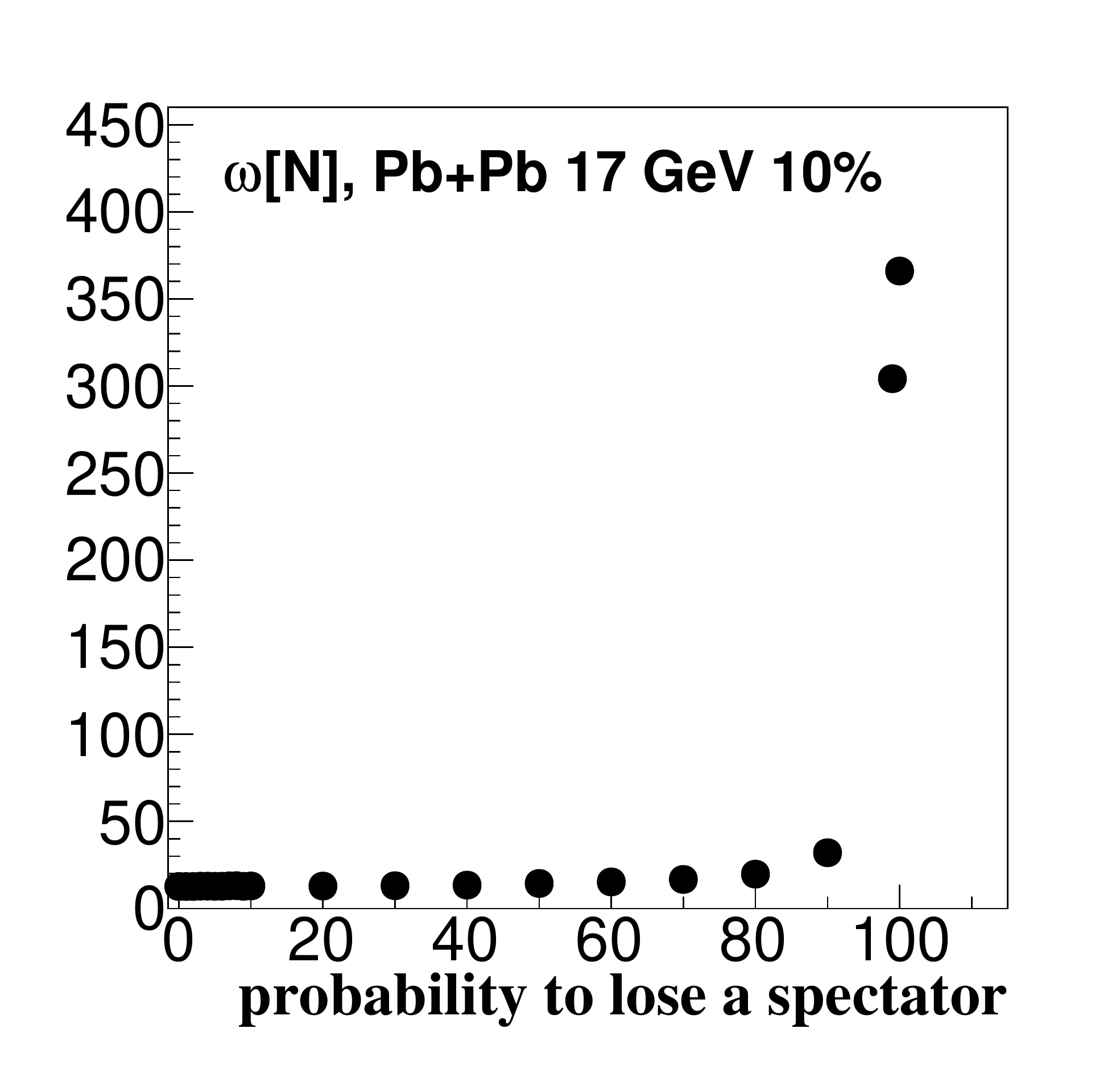}

\caption{Multiplicity and scaled variance versus the probability to lose a forward spectator in WNM with a centrality selection based on the number of detected forward spectators.} \label{fig5}
\end{figure}

\begin{figure}[H]
\centering
\includegraphics[width=6 cm]{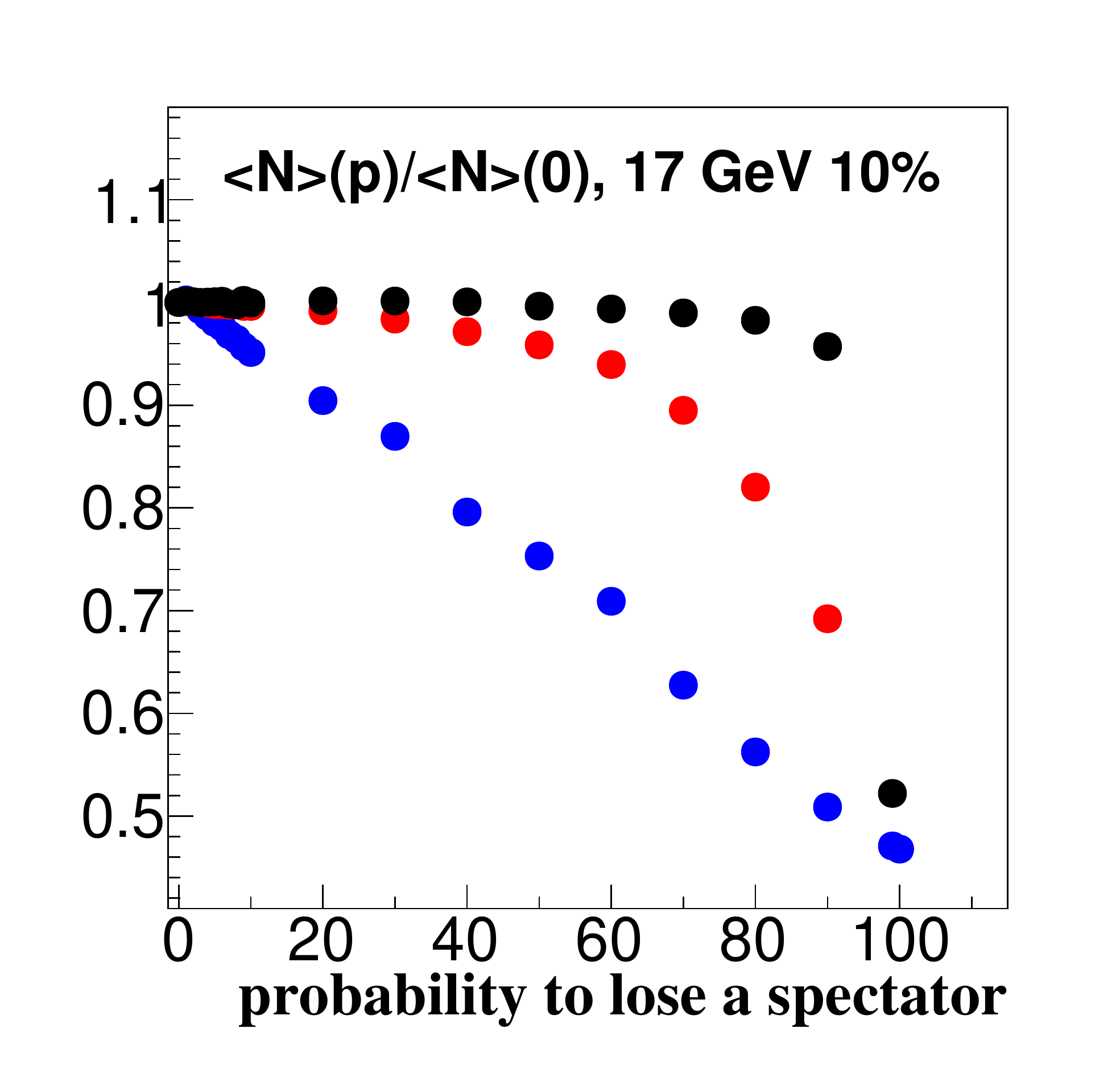}
\includegraphics[width=6 cm]{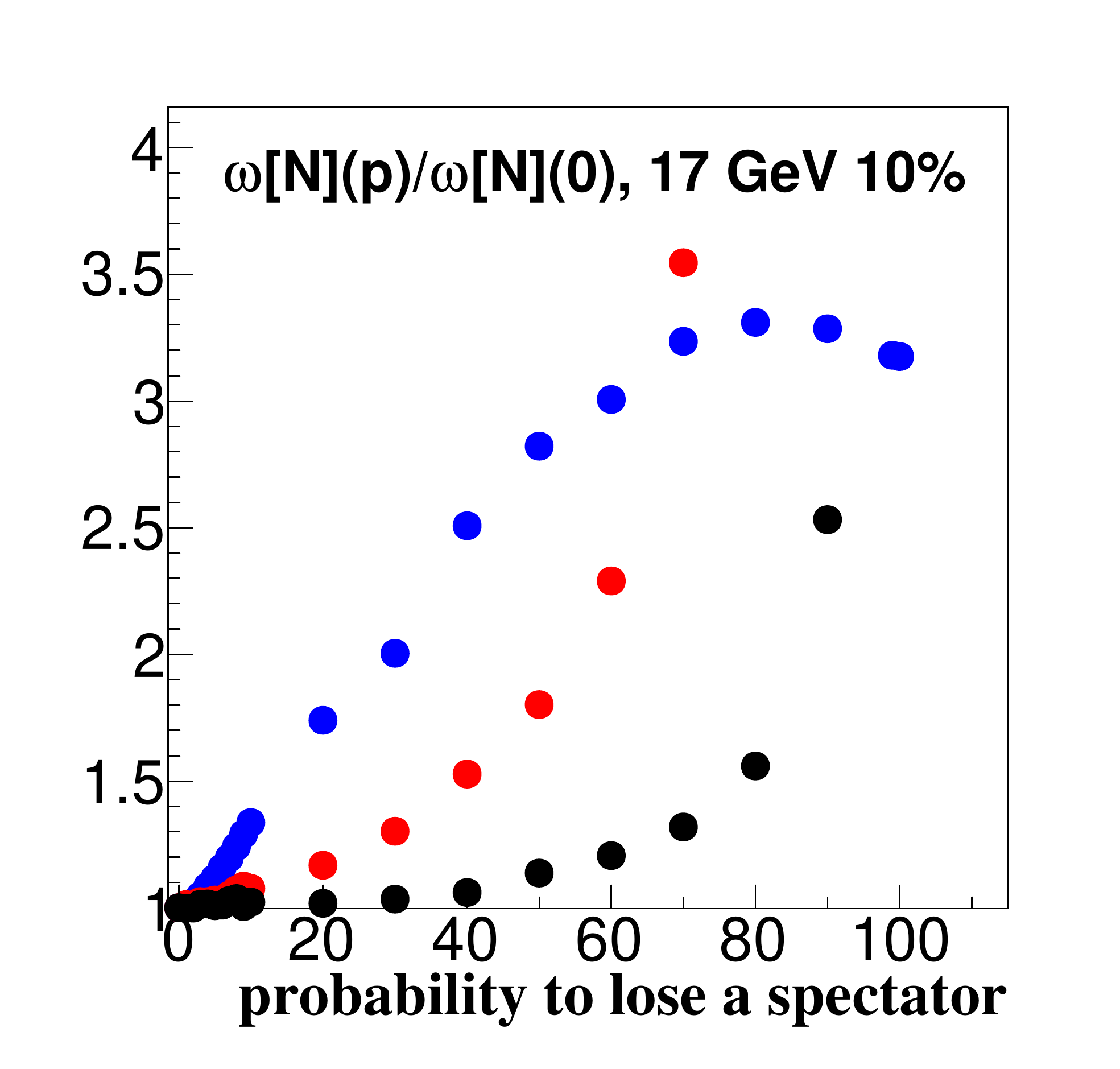}
\caption{Ratios of multiplicity and scaled variance to values with a zero probability to lose a forward spectator for Li + Be (blue), Cl + Ca (red) and Pb + Pb (black) collisions. WNM with a centrality selection based on the number of detected forward spectators.} \label{fig6}
\end{figure}

The dependencies of average event multiplicity and multiplicity scaled variance  $\omega$[N] versus the~probability to lose a forward spectator showed a striking difference {(Figure \ref{fig5})} in the sensitivity~to~detector efficiency between light system and heavy one (${}^7$Li + ${}^9$Be and ${}^{208}$Pb + ${}^{208}$Pb). <N> and $\omega$[N] in Beryllium collisions became sensitive to the spectators lost already, then \emph{p} $\approx$ 3--4\% contrary to Pb + Pb collisions where <N> and $\omega$[N] were steady to the effect until \emph{p} $\approx$ 70\% and 30\% respectively (see  Figure \ref{fig6}).
\vspace{-15pt}

\begin{figure}[H]
\centering
\includegraphics[width=8 cm]{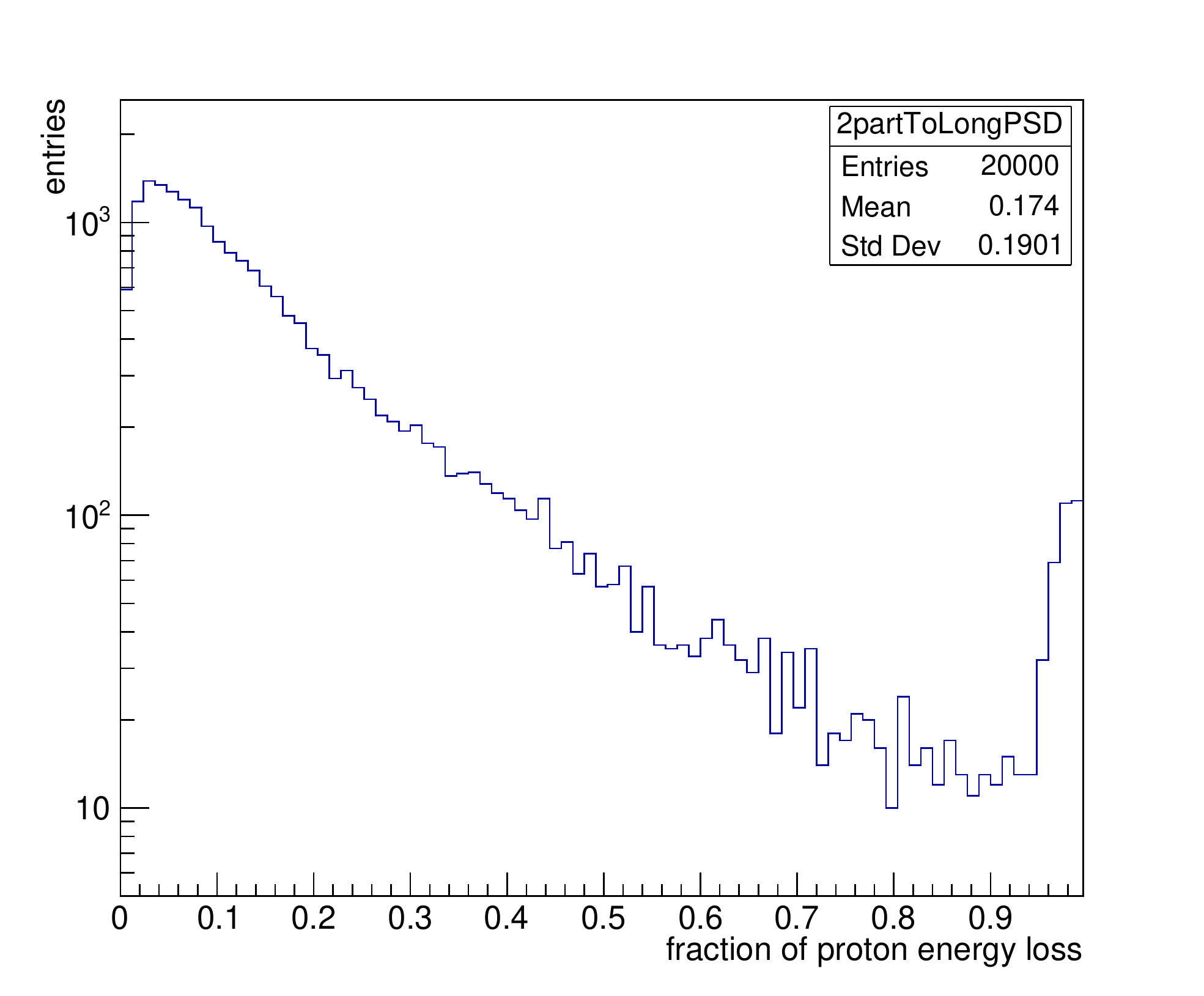}
\caption{A ratio of deposited energy by a 150 GeV/c proton in the sections from 8 to 20 to the whole long calorimeter PSD (11.2 nuclear int. lengths) in the GEANT4 simulation. The whole calorimeter model has 20~sections. This distribution shows the fraction of a proton energy leak from a calorimeter, which has 3.9~nucl. int. lengths or seven sections in case of PSD.} \label{fig7}
\end{figure}

Unexpectedly, this simple model reproduced two important features: higher sensitivity of light systems for small energy loss and lower sensitivity for large fraction energy loss. Nevertheless, we are aware of the fact that the probability to lose a spectator is not a realistic model of a hadronic calorimeter.  The next step was to  introduce a realistic energy losd. For this goal a two-times longer (\begin{math}\approx\end{math}11.2 interaction length) GEANT4 model of PSD was used and a response on a 150 GeV/c proton beam was generated.  We calculated and fitted a distribution of ratio between deposited by a proton energy in the first seven sections ($\approx$3.9 int.l.) to the whole calorimeter (20 sections), as shown in Figure \ref{fig7}. We used the obtained function to introduce a  random energy loss of each forward spectator in the wounded nucleon model.

\vspace{-12pt}

\begin{figure}[H]
\centering
\includegraphics[width=15 cm]{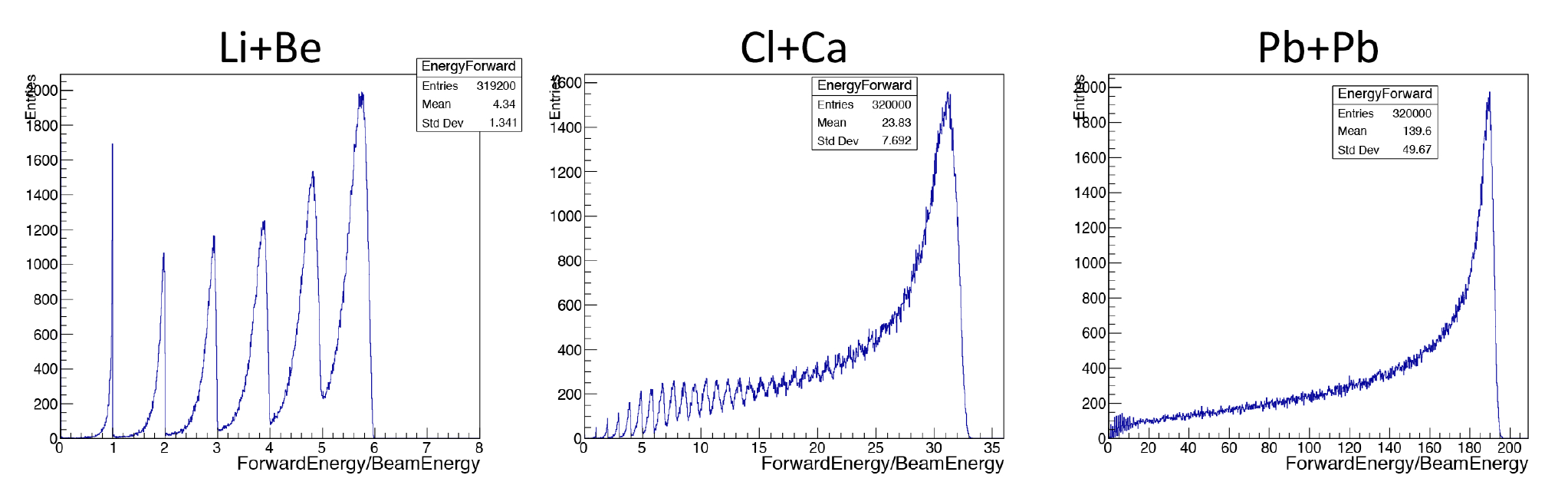}
\caption{Energy distributions of forward spectators with realistic energy leakage from the calorimeter in~WNM. \label{fig8}}
\end{figure}

The 10\% of the most central events were selected by the spectator deposited energy (see  Figure \ref{fig8}). <N> and $\omega$[N] were calculated and compared with the ideal case (without energy loses). The results are presented in   Table   \ref{tab1}.

     \vspace{-12pt}

\begin{table}[H]
\caption{Comparison of results for 10\% most central events with realistic energy leakage and without it in a frame of~WNM.} \label{tab1}
\centering
\begin{tabular}{cccc}
\toprule
\textbf{}	& \textbf{Li + Be}	& \textbf{Ca + Cl}    & \textbf{Pb + Pb}\\
\midrule
<N> without energy loss		& 32.527+/$-$0.017			& 172.069+/$-$0.027		& 1094.34+/$-$0.19 \\
<N> with energy loss		& 32.414+/$-$0.012			& 171.996+/$-$0.034		& 1094.16+/$-$0.15  \\
\begin{math} \omega\end{math}[N]  without energy loss		& 2.0192+/$-$0.0041			& 3.2594+/$-$0.0072 			& 13.101+/$-$0.021	\\
\begin{math} \omega\end{math}[N]  with energy loss		& 2.0625+/$-$0.0042			& 3.3111+/$-$0.0051			& 13.175+/$-$0.019\\
with/without (N)       & 0.99653+/$-$0.00057 & 0.99958+/$-$0.00032 & 0.99984+/$-$0.00052 \\
with/without ($\omega$[N]) &  1.0214+/$-$0.0021 & 1.0159+/$-$0.0023 & 1.006+/$-$0.0017 \\
\bottomrule
\end{tabular}
\end{table}

As shown, the effect is very tiny but nevertheless the lighter the system is the more sensitive it is. The~size of the difference is probably a result of an absence of the energy resolution due to the calorimeter sandwich structure, as   present in the GEANT4 simulation.

\section{Conclusions}

It was observed that the light nuclei systems as Li + Be are more sensitive to the energy leakage from the back side of hadronic calorimeters used for centrality determination in fixed target experiments compared to intermediate size systems as Ar + Sc. The probable reason is that in the light systems most of the forward energy is concentrated only in a few   nucleons. Therefore, a single nucleon loss produces much bigger volume fluctuations than in a collision of heavy systems, which have a presence of more or less constant energy leakage in each collision.  Nevertheless, more investigations are needed to reach the complete understanding of the phenomenon. Even though we succeeded in demonstrating the sensitivity of light systems in the framework of the wounded nucleon model with the probability of a spectator loss, the realistic energy loss simulation in the same model shows only a tiny effect on average multiplicity and scaled variance.

The future fixed target programs, which aim to study light nuclei colliding systems, have to pay attention that a longer calorimeter is needed to control the volume fluctuations for such reactions than for heavier ones.

\vspace{6pt}
\funding{This research was funded by the Russian Science Foundation grant number 17-72-20045 in a part of data analysis from MC generators (Section \ref{sec2}) and the Russian Foundation for Basic Research grant number 18-32-01055 mol\_a in a part of wounded nucleon model analysis (Section \ref{sec3}).}

\acknowledgments{We would like to thank Sergey Morozov, Marina Golubeva and the PSD group for providing the calorimeter and the SHIELD MC simulations; Justyna Monika Cybowska for providing the EPOS MC simulation; and~Evgeny Andronov, Marek Gazdzicki, Peter Seyboth and the NA61/SHINE collaboration for discussions leading to this article.}

\conflictsofinterest{The author declares no conflict of interest. The funders had no role in the design of the study; in the collection, analyses, or interpretation of data; in the writing of the manuscript, or in the decision to publish the~results.}


\reftitle{References}




\end{document}